# PENETRATION TESTING IN AGILE SOFTWARE DEVELOPMENT PROJECTS

Martin Tomanek and Tomas Klima

Department of Systems Analysis, University of Economics, Prague, Czech Republic

## ABSTRACT

*Agile development methods are commonly used to iteratively develop the information systems and they can easily handle ever-changing business requirements. Scrum is one of the most popular agile software development frameworks. The popularity is caused by the simplified process framework and its focus on teamwork. The objective of Scrum is to deliver working software and demonstrate it to the customer faster and more frequent during the software development project. However the security requirements for the developing information systems have often a low priority. This requirements prioritization issue results in the situations where the solution meets all the business requirements but it is vulnerable to potential security threats.*

*The major benefit of the Scrum framework is the iterative development approach and the opportunity to automate penetration tests. Therefore the security vulnerabilities can be discovered and solved more often which will positively contribute to the overall information system protection against potential hackers.*
*In this research paper the authors propose how the agile software development framework Scrum can be enriched by considering the penetration tests and related security requirements during the software development lifecycle. Authors apply in this paper the knowledge and expertise from their previous work focused on development of the new information system penetration tests methodology PETA with focus on using COBIT 4.1 as the framework for management of these tests, and on previous work focused on tailoring the project management framework PRINCE2 with Scrum.*

*The outcomes of this paper can be used primarily by the security managers, users, developers and auditors. The security managers may benefit from the iterative software development approach and penetration tests automation. The developers and users will better understand the importance of the penetration tests and they will learn how to effectively embed the tests into the agile development lifecycle. Last but not least the auditors may use the outcomes of this paper as recommendations for companies struggling with penetrations testing embedded in the agile software development process.*

## 1. INTRODUCTION AND METHODOLOGY

In 1970 the waterfall development model was introduced by Winston W. Royce. Since this time the model has been used to manage majority of software development projects. This model follows a phased approach in which the requirements are defined upfront then the solution is designed, coded, tested and released to production. In many cases this approach was successful but unfortunately in many cases failed. The chaos report from the year 2013 [1] that focuses on project success rate indicates that 39% of projects were delivered successfully, 18% completely failed and 43% were challenged. The project success rate was slightly improved compared to year 2004 but the percentage of failed projects still remains almost the same.

The current business environment demands the higher success rate and shorter time to market and greater flexibility of the ever-changing business requirements [2], [3]. Many IT leaders in the software development field found out that robust and document-oriented process frameworks simply do not work as expected. They introduced the agile manifesto [4] which is followed by many software developers around the world. As a result of introducing the agile manifesto, several agile development frameworks were introduced with different scope and focus (e.g. Scrum, Extreme Programming, FDD etc.).

A shift to agile methods can increase the success rate and mitigate some issues that are typical for traditional phased-oriented frameworks [5]. Agile development methods can be used not only for developing small and simple software but they are suitable for development of big and complex IT systems. Good examples of the successful usage of the agile frameworks on large software development projects can be found in the case studies [6], [7].

The most widely used agile frameworks are Scrum and the hybrid framework combining Scrum and Extreme programming framework [8]. The framework in the scope of this paper is Scrum. Scrum was developed by Schwaber and Sutherland and is described in the Scrum Guide [9].
Penetration testing represents the comprehensive approach to identify the real information systems vulnerabilities that can be exploited by hackers. Penetration tests are executed by so called ethical hackers who use their knowledge and experience to break the overall security measures. All these found vulnerabilities are then identified and necessary actions proposed in order to strengthen the information systems security. Some penetration tests can be also automated and executed on a regular basis that consistently ensures that information systems meet some level of security assurance.

One of the advantages of the penetration tests is related to seniority of ethical hackers who can follow the most up-to-date identified threats in IT related technologies and can protect the company's assets against the external hackers. The security threats are evolving very fast and it is difficult to reflect them in the current security standards and guidelines that are world-wide used to secure the information systems. The companies realize this trend in IT security field and put the penetration tests in one of the four most important aspects of current IT security management [10].

The biggest challenges the companies faced to, in regard to penetration tests, is how to embed the penetration testing into the overall IT management framework. Penetration tests can be executed on overall information systems landscape including the IT infrastructure or they can be embedded into the software development process.

The goal of this paper is to introduce a way how to embed the penetration testing into the Scrum framework that represents the most used agile software development framework. The benefit of this paper is to propose how Scrum can help to automate the penetration tests during the software development projects, incorporate the specialized penetration tests into the regular software releases and improve the overall resistance of developing software.

Advantages and disadvantages of the agile software development framework with included penetration tests will be discussed as well. This contribution is conceived as design research, the result of which is an artefact.

## 2. SCRUM

Scrum is an agile software development framework that is mainly used for iterative and incremental software development. The core of the Scrum framework is that customer requirements can be changed during development and software is developed iteratively. Iterations are called sprints and every sprint starts with a sprint planning meeting where the customer reviews and prioritizes requirements. The requirements are recorded as user stories that represent the requirements from the customer perspective. All user stories are stored in the product backlog. The prioritized requirements that can be delivered by the development team are selected, agreed and transferred to sprint backlog that is used to manage the requirements during the sprint. Then the development team works together to develop software features, satisfy customer requirements and deliver shippable software by the end of each sprint. This shippable software increment is presented in the sprint review meeting where the customer can it and think about further requirements. The most frequent meetings are daily stand-up meetings where development team members discuss what they have done since the last meeting, what they will do in the coming days and whether they are facing any impediments.

The Scrum framework is depicted on the following picture that graphically illustrates the agile software development process.

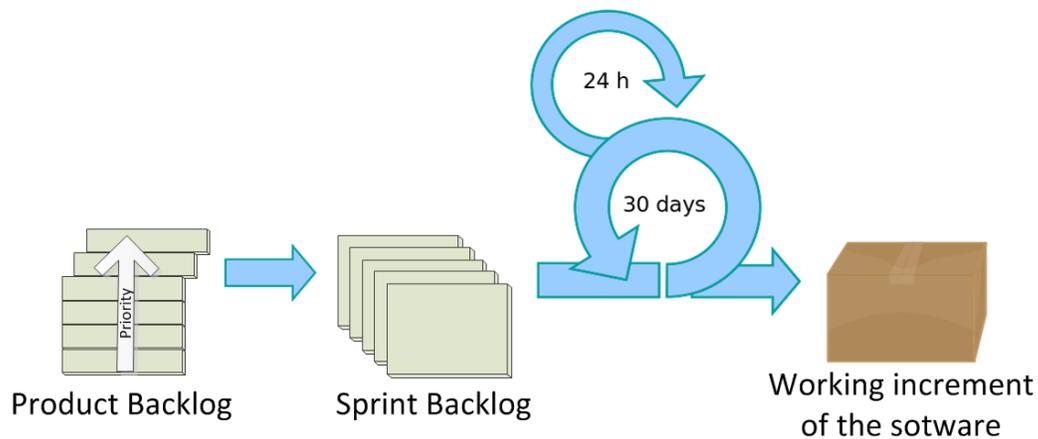

Figure 1.  Scrum framework. Source: authors based on [9]

Three roles are defined in the Scrum framework. The Product Owner represents customers and is responsible for defining and prioritizing software requirements and records them in the product backlog. The Development Team is responsible for delivering the potentially shippable software by the end of each sprint. The Scrum Master facilitates Scrum meetings and ensures the development team can work as efficient as possible.

Scrum is the product-oriented framework that focuses mainly on team cooperation but also on quality of each product increment. The concept of "Done" definition is introduced. There should be a common understanding of all requirements that should be completed by the end of every sprint so the increment is working and can be released to production. These requirements usually cover all the necessary tests, delivery of required documents and other. When all these requirements are completed then the software increments can be considered as "Done".

## 3. EMBEDDING PENETRATION TESTS INTO SCRUM

Now the question is how to embed the penetration tests into Scrum. As mentioned in the introduction chapter, there are two types of penetration tests: the automated and manual penetration tests.

The real tangible benefit of using Scrum, which is mentioned by various researchers, is an opportunity to automate the tests in iterations and fix the defects and vulnerabilities earlier and on a regular basis [11]–[14]. Therefore Scrum provides the solid foundation for automating the penetration tests in agile software development process.

As the software is developed iteratively piece by piece, it doesn't make sense implement and test all security requirements in the first sprint. Therefore authors of this paper suggest filling the product backlog with security and penetration test requirements at the earliest stage of the development and continuously prioritize these requirements and select the most value-adding or risky requirements to be developed first and continuously tested in upcoming sprints. The designed and implemented automated tests can be then executed regularly in every sprint and lower the costs of tests execution and later debugging.

The manual penetration tests are usually executed by ethical hackers on the existing running information systems. Therefore the authors suggest including the manual penetration tests not in every sprint but only if the product owner decides to release the software into production. Of course if the release contains only minor changes then it won't make sense to involve the specialized ethical hacker to test it. However if the release contains major or complex changes then the manual penetration tests should be executed in order to accredit the IT solution. The decision about the manual penetration tests doesn't depend only on the complexity of changes but it depends on more factors like: does the software contain confidential or financial data, how critical is the software to run the business, and is the software accessible publicly from internet?

The Scrum framework with security requirements and penetration tests is depicted on the following picture. The product backlog is enriched by security requirements that can come from the security standard like ISO27002, the more generic framework like COBIT 5.0 and of course from the experienced ethical hackers. Automated and manual penetration tests are added as well to illustrate the placement of tests.

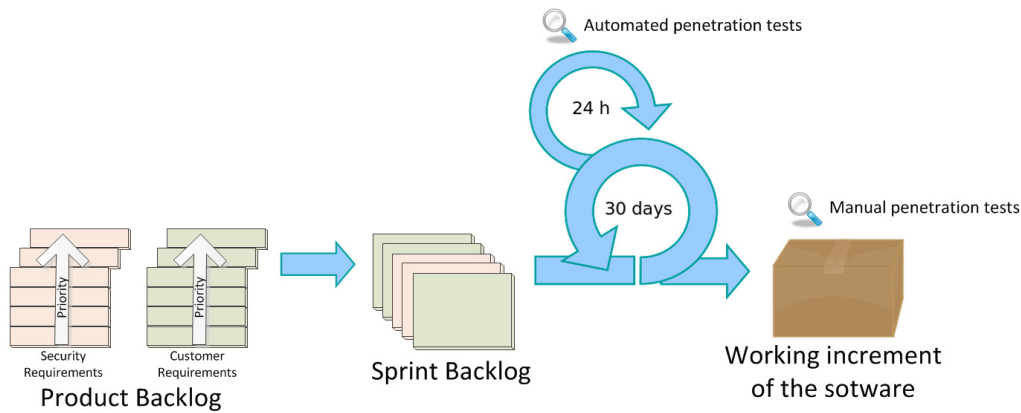

Figure 2: Scrum framework with embedded penetration tests. Source: authors

Last but not least the penetration testing should be also reflected in the definition of "Done". In this case the requirements can be formulated as follows:

- Automated penetration tests have been executed on the final version of the software increment
- All critical vulnerabilities or defects resulting from the penetration tests have been captured and fixed
- Not fixed minor vulnerabilities or defects have been evaluated and recorded to the product backlog for further consideration

## 4. CONCLUSION AND FURTHER RESEARCH

This paper introduces the enriched Scrum agile software development framework that includes the definition of security requirements and execution of automated and manual penetration tests. This concept has been introduced and validated in various software development projects in the global logistics company. The following advantages and disadvantages are formulated based on lessons learnt of individual software development projects.

The major advantage, of the enriched Scrum framework by penetration tests, the authors of this paper see in the inclusion of regular automated tests in every iteration cycle of software development that result in:

- Regular and more frequent security and penetration tests
- Improved efficiency and lower costs of test execution due to test automation
- Earlier defect or security vulnerability detection
- Software vulnerability fixing in early stages of software development and not at the end of software development phase as suggested by waterfall development frameworks
- Measurement of security defects per iteration and opportunity to improve this measurement over the time
- Enforcement of security best practices in software development projects

However also some disadvantages or risks related to the Scrum framework with embedded penetration tests were noticed, for example:

- Initial setup of automated penetration tests in the first sprint requires a significant effort and it may result in the limited functionality of the first version of the working increment of the software
- Increased development costs due to licenses of specialized test automation tools
- Increased complexity of software development tools

The further authors' research is focused on development of the penetration testing methodology PETA. This methodology combines current trends in information systems security area with well sound best practices like COBIT 4.1, COBIT 5.0, PRINCE2 and others. In regards to agile software development process, the usage of PETA methodology will result in identification of security vulnerabilities in the company IT landscape that will feed the security requirements in the product backlog.

## ACKNOWLEDGEMENTS

This paper was prepared thanks to the IGA grant VSE IGS F4/5/2013.

## AUTHORS

**Martin Tomanek**

Martin Tomanek graduated from applied informatics at the Faculty of Informatics and Statistics, University of Economics, Prague. Currently, he is PhD student at the Department of Systems Analysis, Faculty of Informatics and Statistics, University of Economics, Prague, where he develops the integrated framework based on PRINCE2, Scrum and other best practices used in SW development area.

**Tomas Klima**

Tomas Klima graduated from applied informatics at the Faculty of Informatics and Statistics, University of Economics, Prague. Currently, he is PhD student at the Department of Systems Analysis, Faculty of Informatics and Statistics, University of Economics, Prague, where he deals with information security and develops a new framework for penetration testing.